\documentclass[a4paper]{article}

\usepackage{INTERSPEECH2022}
\usepackage{float}
\newtheorem{prop}{Proposition}
\newtheorem{theorem}{Theorem}
\usepackage{booktabs}
\usepackage{hyperref}

\title{Complex Frequency Domain Linear Prediction: \\
A Tool to Compute Modulation Spectrum of Speech}
\name{Samik Sadhu$^1$, Hynek Hermansky$^{1,2}$}
\address{
  $^1$Center for Language and Speech Processing, Johns Hopkins University, USA\\
  $^2$Human Language Technology Center of Excellence, Johns Hopkins University, USA}
\email{samiksadhu@jhu.edu, hynek@jhu.edu}

\begin{document}

\maketitle
\begin{abstract}
 Conventional Frequency Domain Linear Prediction (FDLP) technique models the squared Hilbert envelope of speech with varied degrees of approximation which can be sampled at the required frame rate and used as features for Automatic Speech Recognition (ASR). Although previously the complex cepstrum of the conventional FDLP model has been used as compact frame-wise speech features, it has lacked interpretability in the context of the Hilbert envelope. In this paper, we propose a modification of the conventional FDLP model that allows easy interpretability of the complex cepstrum as temporal modulations in an all-pole model approximation of the power of the speech signal. Additionally, our ``complex" FDLP yields significant speed-ups in comparison to conventional FDLP for the same degree of approximation. 
\end{abstract}


\section{Introduction}
\textit{``Message is carried in changes in vocal tract shape, which modulate spectral components of speech" - Homer Dudley}. \\

Human beings have learnt to communicate by modulating the power of speech signals, predominantly below 30 Hz with an energy peak around 4-5 Hz \cite{ding2017temporal,hermansky1997modulation,greenberg1996insights}. Concurrently, human hearing has evolved to maintain a symbiotic relationship with human speech production. In prior studies, human listeners reported severe reduction in intelligibility when important speech modulations were selectively deleted \cite{kanedera1997importance,drullman1994effect}. Not surprisingly, related studies on machine recognition of speech has exhibited similar results \cite{kanedera1997importance,arai1996intelligibility}. 

Traditionally, speech modulations have been obtained by band-pass filtering the time trajectories of the power of speech signal in different frequency sub-bands \cite{hermansky1997modulation,drullman1994effect,598826,drullman1994}. Eventual development of tools to mathematically model the power trajectories of speech signal such as Frequency Domain Linear Prediction (FDLP)  led to the use of the \textit{complex cepstrum} of the FDLP model as compact frame-wise representation of speech \cite{athineos2004lp,1318451}. A connection between \textit{complex cepstrum} and the \textit{modulation spectrum}, although intuitively obvious, remains to be established. In this work, we formally establish this connection, show its fundamental limitation when interpreted as temporal modulations and propose a complex valued FDLP modeling of the power of speech signal to resolve this limitation. 

\section{Frequency Domain Linear Prediction (FDLP)}
Linear Prediction (LP) is used to perform discrete time-series analysis where the current signal value is predicted as a weighted combination of its past values \cite{makhoul1975linear} as in Eq. \ref{eq:lp}
\begin{equation}
\label{eq:lp}
    x[n]=\sum_{k=1}^p\alpha_kx[n-k]+Gu[n]
\end{equation}
, where $x$ is the discrete time signal, $n$ denotes discrete time points, $p$ is the model order, $G$ is the model gain and $u$ is the input to the LP model. 

\begin{figure}[H]
    \centering
    \includegraphics[scale=0.32]{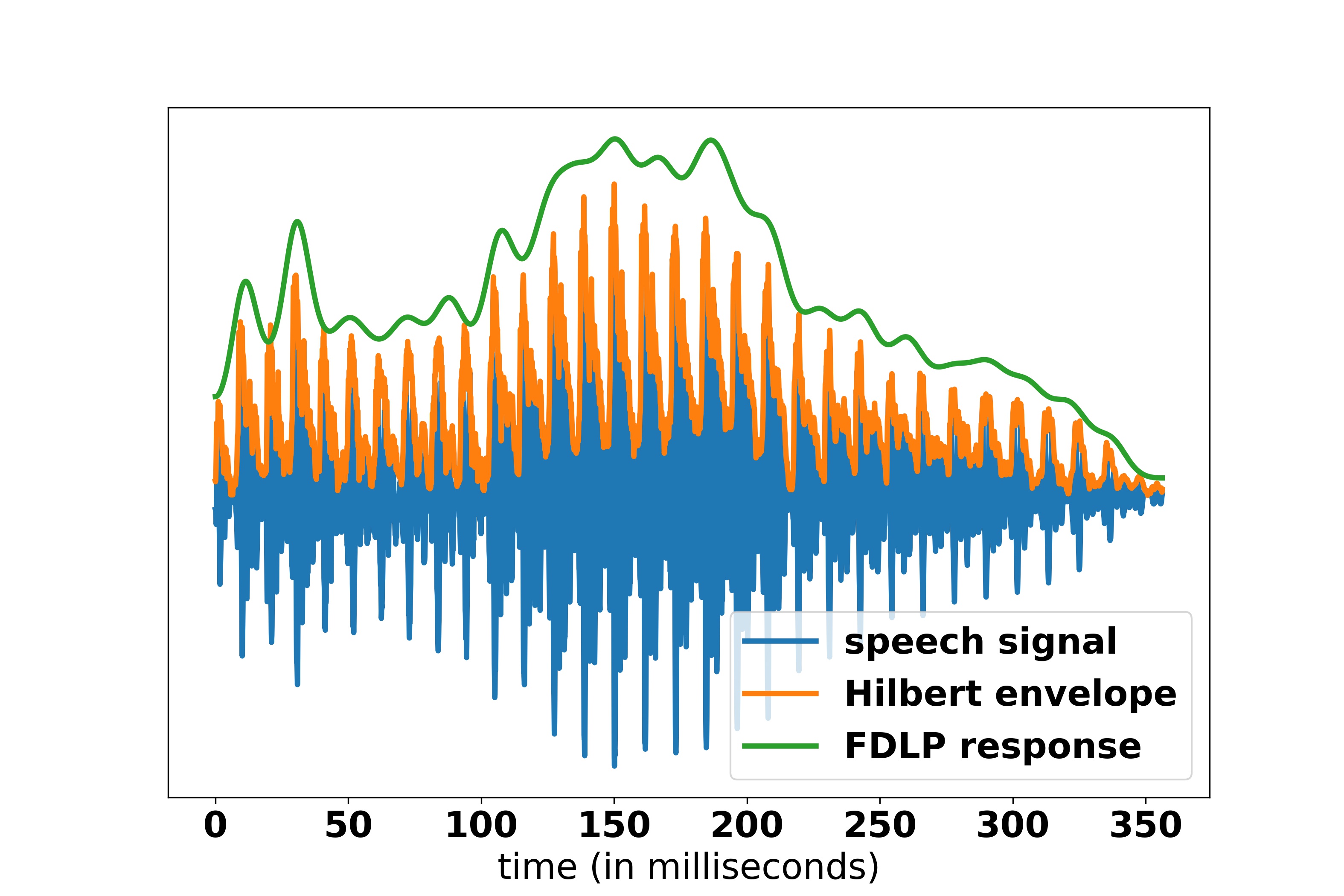}
    \caption{A sample speech waveform of the word ``hello" and its Hilbert envelope are shown, together with half of the FDLP fit to the even symmetric squared Hilbert envelope with model order 50.}
    \label{fig:hilbert_envelope}
\end{figure}

One of the foremost properties of LP analysis of a signal is that it results in an all-pole transfer function whose frequency response  fits the power spectrum of the original signal to varying levels of approximation based on the filter order. An analogous application of linear prediction was used by Marios et al. \cite{athineos2004lp,athenaos_thesis} where this property of linear prediction was utilized in a dual fashion resulting in Theorem \ref{Th:1}.
\begin{theorem}
\label{Th:1}
Linear prediction of the Discrete Cosine Transform (DCT) of a signal results in an all-pole transfer function whose time response approximates the even-symmetrized squared Hilbert envelope of the signal. 
\end{theorem}
A proof of this statement can be found in  \cite{athenaos_thesis}. Hilbert envelope is the magnitude of the analytic signal of the original signal and, as the name suggests, is one of several ways to compute the envelope of a signal (Figure \ref{fig:hilbert_envelope}). For the remainder of the paper, we will be referring to this technique as \textit{conventional} FDLP.

\subsection{Modulation Spectrum from Conventional FDLP}
Temporal modulations of a signal are the frequency components in its time envelope. Therefore, the frequency components in the all-pole approximation of the Hilbert envelope after any acceptable transformation is a logical definition of modulation spectrum obtained from a FDLP model. Since conventional FDLP fits the \textit{even-symmetrized} Hilbert envelope, for computing modulations by direct Fourier transform of the FDLP response we need to only preserve the first half of the all-pole approximation.


Computing the modulation spectrum in the aforementioned way, requires two fast Fourier transforms - firstly to compute the time response of the all-pole transfer function and secondly to compute the modulation spectrum. Each transform requires $\mathcal{O}(N\log N)$ computations, N being the number of data points in the original signal and thus are computationally expensive operations. 

Marios et al. \cite{athineos2004lp, 1318451} used the complex cepstrum \cite{oppenheim1968homomorphic} of the FDLP model as computationally superior and compact features from the FDLP model - a connection to the modulation spectrum was suggested but not elucidated. Considering $\mathcal{F}$ to be the discrete time Fourier transform, the complex cepstrum of an FDLP model is defined as 
\begin{eqnarray}
    c[f]&=& \mathcal{F}^{-1}\log H(e^{j\tau}) 
\end{eqnarray}
, where $f$ denotes discrete frequency points determined by the original signal length $T$ and $\tau=\frac{2\pi  t}{T}$ represents the ``normalized radian time" where $t$ represents the original un-normalized time. As a result, $\tau=0$ and $\tau=2\pi$ normalized radian time denotes the beginning and end of the even-symmetric signal respectively. $H(e^{j\tau})$ is the time response of the FDLP all-pole transfer function. There exists a computationally efficient way to compute the complex cepstrum directly from the all-pole model coefficients $\{\alpha_k, k=1,2, \dots p \}$ \cite{sadhu2021radically}. 

Practically, we can only compute the Discrete Fourier Transform (DFT) of a signal on a digital device, for which $|\mathcal{F}^{-1}\log H(e^{j\tau})|=|\mathcal{F}\log H(e^{j\tau})|$ (see Appendix A). This bridges the gap between complex cepstrum and modulation spectrum and leads to a judicious proposition to compute the modulation spectrum from a conventional FDLP model using the cepstral recursion. 

\begin{prop}
\label{prop:1}
The magnitude of the complex cepstrum $|c[f]|$ of a conventional FDLP model obtained from a signal gives the modulation spectrum of the even-symmetrized signal. 
\end{prop}

Proposition 1 justifies the use of the complex cepstrum as ASR features in \cite{athineos2004lp}, and eventually in the definition of modulation vectors in \cite{sadhu2019m,sadhu2019modulation} to compute sub-band based robust modulation features for ASR.
\subsubsection{The Main drawback of Proposition \ref{prop:1}}
\begin{figure}[H]
    \centering
    \includegraphics[scale=0.35]{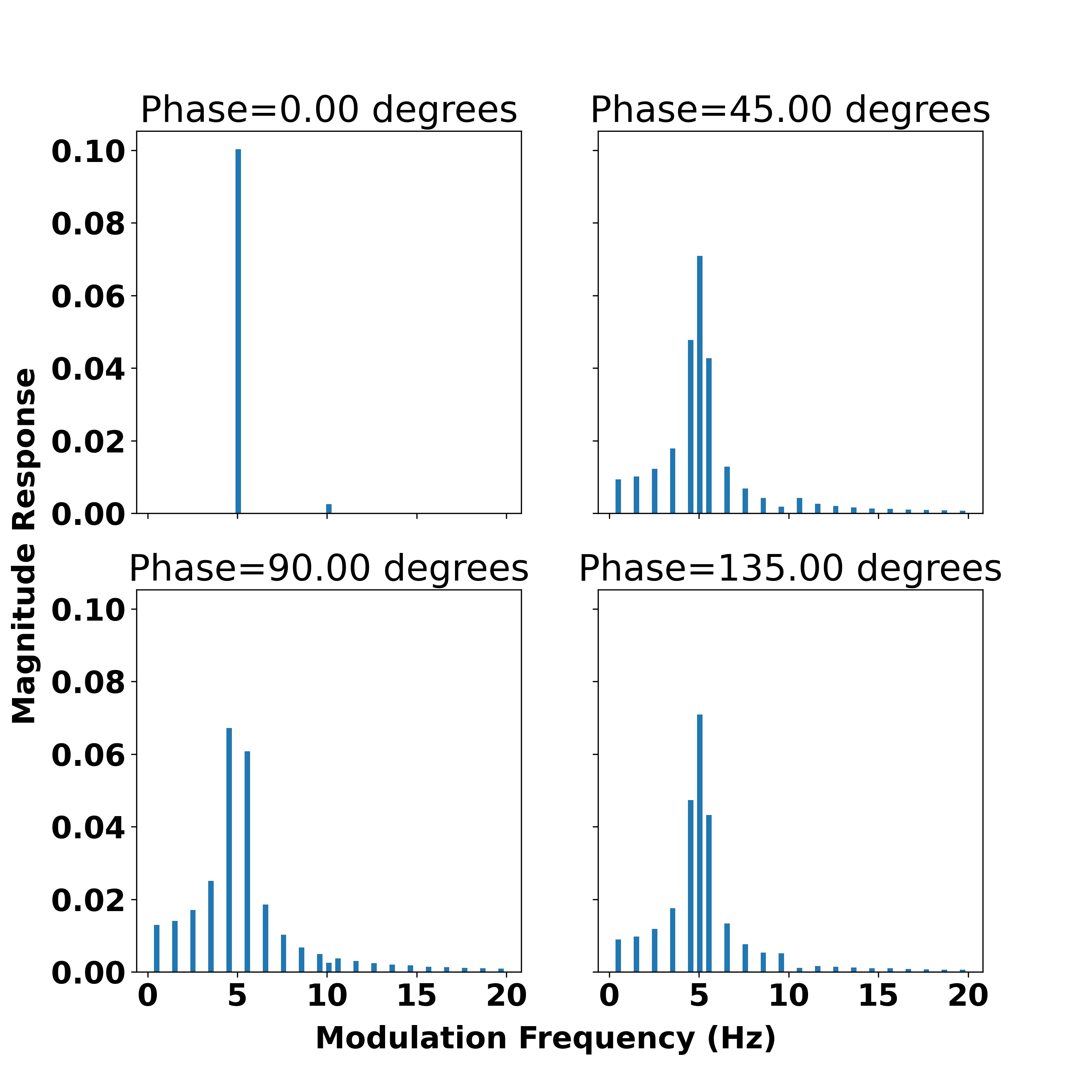}
    \caption{$|c[f]|$ for conventional FDLP models obtained from a 1 second long modulated signal $x(t)$ with $f^{(c)}=1000$ Hz, $f^{(m)}=5$ Hz and modulation depth $\gamma=0.1$. The modulation phase shift $\phi$ takes four different values. The correct modulation spectrum is obtained only for $\phi=0$ degrees. When $\phi=90$ degrees, the 5 Hz modulation is not detected at all.}
    \label{fig:phase}
\end{figure}

Consider an amplitude modulated signal 
\begin{equation}
    x(t)=\left(1-\gamma\cos(2\pi f^{(m)}t+\phi)\right)\sin(2\pi f^{(c)}t)
\end{equation}
with carrier frequency $f^{(c)}$ Hz, modulation frequency  $f^{(m)}$ Hz, modulation depth $\gamma$ and modulation phase $\phi$.
Unlike the direct computation of modulation from half of the FDLP response, $|c[f]|$ captures modulations in the even symmetrized FDLP response, making them real valued numbers in the DCT domain and sensitive to the phase of the modulation as can be seen in Figure \ref{fig:phase}. 

This results in  ambiguity in the meaning of each of these modulation coefficients. Although a DCT contains the same information as Fourier transform, the representation of a 5 Hz modulation, as shown in Figure \ref{fig:phase}, is not unique.
\section{Complex Frequency Domain Linear Prediction}

A sensible solution to this problem is to obtain all-pole approximations of envelopes of the original signal instead of the even-symmetrized signal. We revisit the property that linear prediction of a signal gives an all-pole transfer function whose frequency response fits the power spectrum of the signal and consider Theorem  \ref{Th:2}. 
\begin{theorem}[Complex FDLP]
\label{Th:2}
Linear prediction of the inverse Fourier transform of a signal results in an all-pole transfer function whose time response approximates the
the power of the signal.
\end{theorem}
A proof of this theorem is straight-forward since the power spectrum of $\mathcal{F}^{-1}(x[n])$ is given by $|\mathcal{F}(\mathcal{F}^{-1}x[n])|^2=x[n]^2$.

\begin{figure}[H]
    \centering
    \includegraphics[scale=0.32]{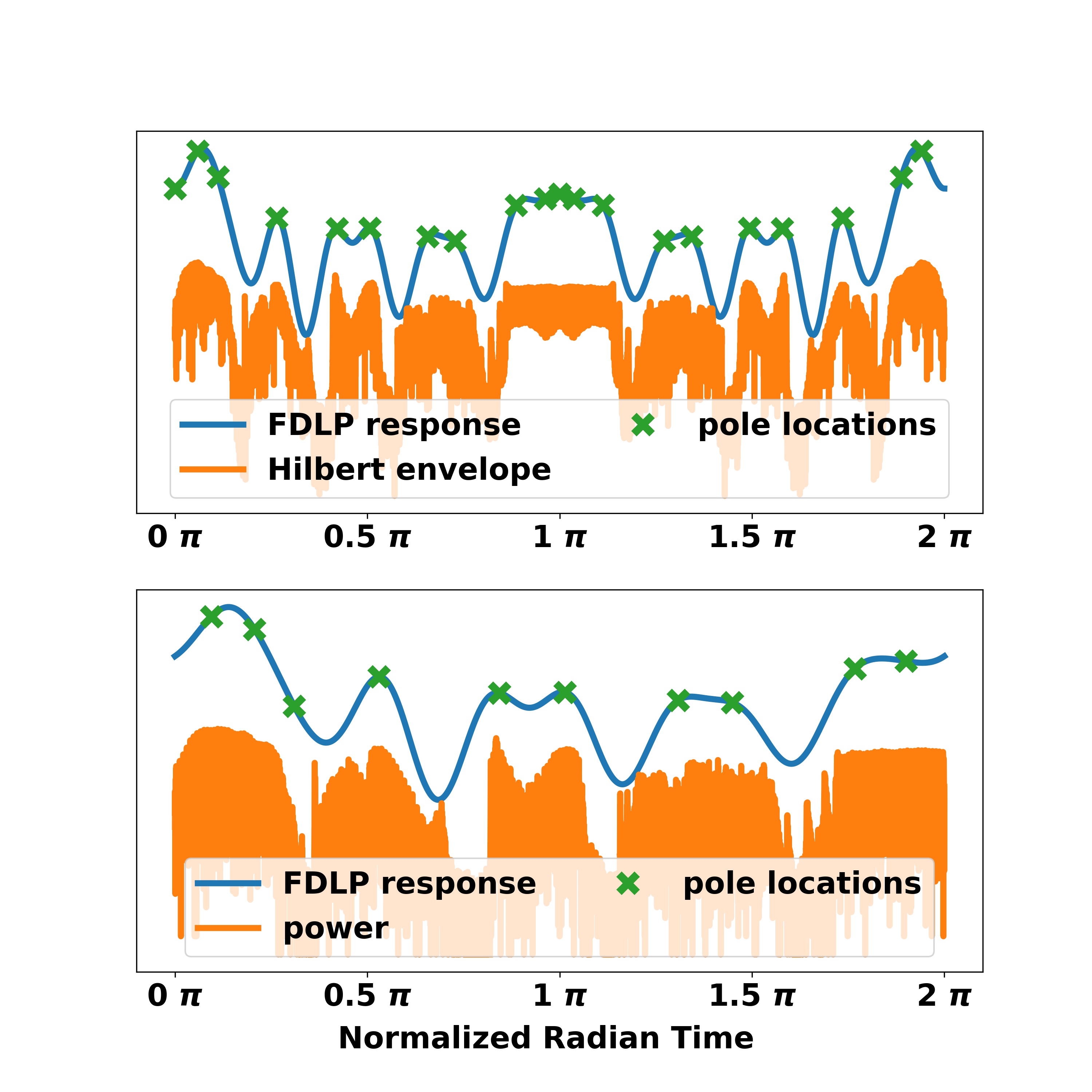}
    \caption{  (Top) Conventional FDLP model fit to the even-symmetric Hilbert envelope of a one second sample of speech with model order 20. The poles can be seen to be conjugate symmetric, i.e. every pole at $\tau=\theta$ has a conjugate pole at $\tau=2\pi-\theta$ \\ (Bottom) Complex FDLP fit to the power of the same one second speech signal with model order 10. The poles do not have conjugate symmetry.}
    \label{fig:compare_fits}
\end{figure}

\noindent The primary differences between conventional FDLP and complex FDLP are 
\begin{itemize}
    \item Complex FDLP, as the name suggests, executes linear prediction with a complex valued signal $\mathcal{F}^{-1}(x[n])$ leading to a complex transfer function
    \item From properties of polynomials, the poles of conventional FDLP are conjugate symmetric which results in the even-symmetric time response. The poles of the complex FDLP filter are not conjugate symmetric and $\tau=0$, $\tau=2\pi$ denote the start and end of the original non-tampered signal ( see Figure \ref{fig:compare_fits} ). 
    \item Unlike the conventional FDLP, the complex cepstrum of the complex FDLP model is complex valued and since $|\mathcal{F}^{-1}\log H(e^{j\tau})|=|\mathcal{F}\log H(e^{j\tau})|$, the magnitude of the modulations can be explicitly computed.
    \item In addition, since the complex FDLP fits the original non-symmetric signal, $|\mathcal{F}\log H(e^{j\tau})|$ is insensitive to the phase of the modulation.
\end{itemize}

Proposition \ref{prop:2} follows.  

\begin{prop}
\label{prop:2}
The magnitude of the complex cepstrum $|c[f]|$ of a complex FDLP model obtained from a signal gives the modulation spectrum of the signal. 
\end{prop}

\begin{figure}[h]
    \centering
    \includegraphics[scale=0.32]{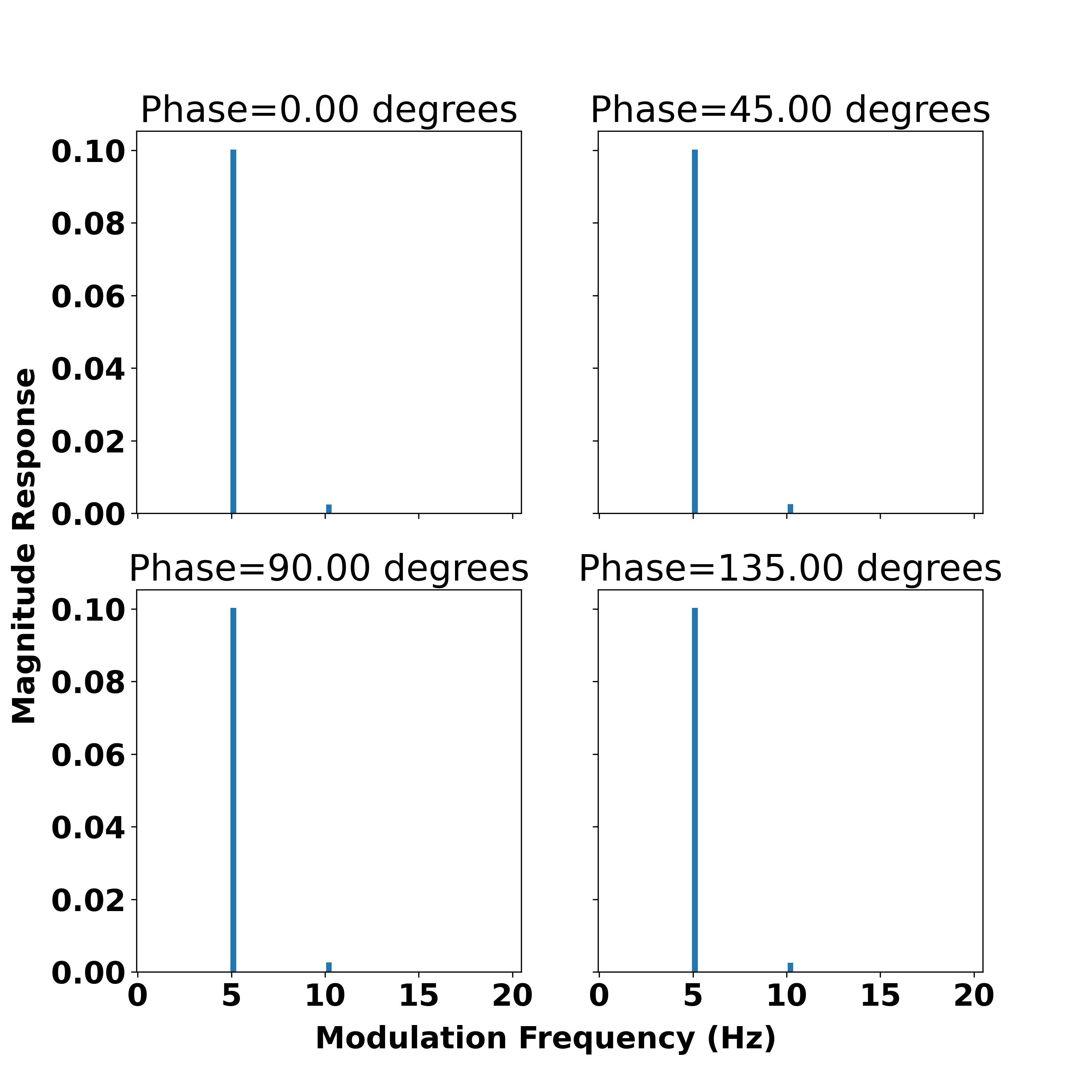}
    \caption{$|c[f]|$ for complex FDLP models obtained from a 1 second long modulated signal $x(t)$ with $f^{(c)}=1000$ Hz, $f^{(m)}=5$ and modulation depth $\gamma=0.1$. The modulation phase shift $\phi$ takes four different values. Complex FDLP is capable of capturing the true magnitude of the modulations at all phase shifts.}
    \label{fig:phase_insensitive}
\end{figure}

Unlike conventional FDLP, figure \ref{fig:phase_insensitive} shows that $|c[f]|$ from complex FDLP model captures the value of the correct modulation and is insensitive to the phase of modulation\footnote{A Python implementation of complex FDLP based modulation spectrum is available at \url{https://github.com/sadhusamik/fdlp_spectrogram}.}.

\subsection{Combination of Modulations}
Although our technique is capable of capturing isolated modulations, applications would require accurate computation of the entire modulation spectrum of a signal. Consider a signal with co-occurring modulations
\begin{equation}
    x(t)=\left(1-\sum_{i=1}^N\gamma_i\cos(2\pi f^{(m)}_it+\phi_i)\right)\sin(2\pi f^{(c)}t)
\end{equation}
with carrier frequency $f^{(c)}$ Hz, modulation frequencies  $f^{(m)}_1,f^{(m)}_2,\dots, f^{(m)}_N $ Hz with modulation depths $\gamma_1,\gamma_2,\dots , \gamma_N $ and phases $\phi_1,\phi_2,\dots, \phi_N$ respectively.
\begin{figure}[h]
    \centering
    \includegraphics[scale=0.33]{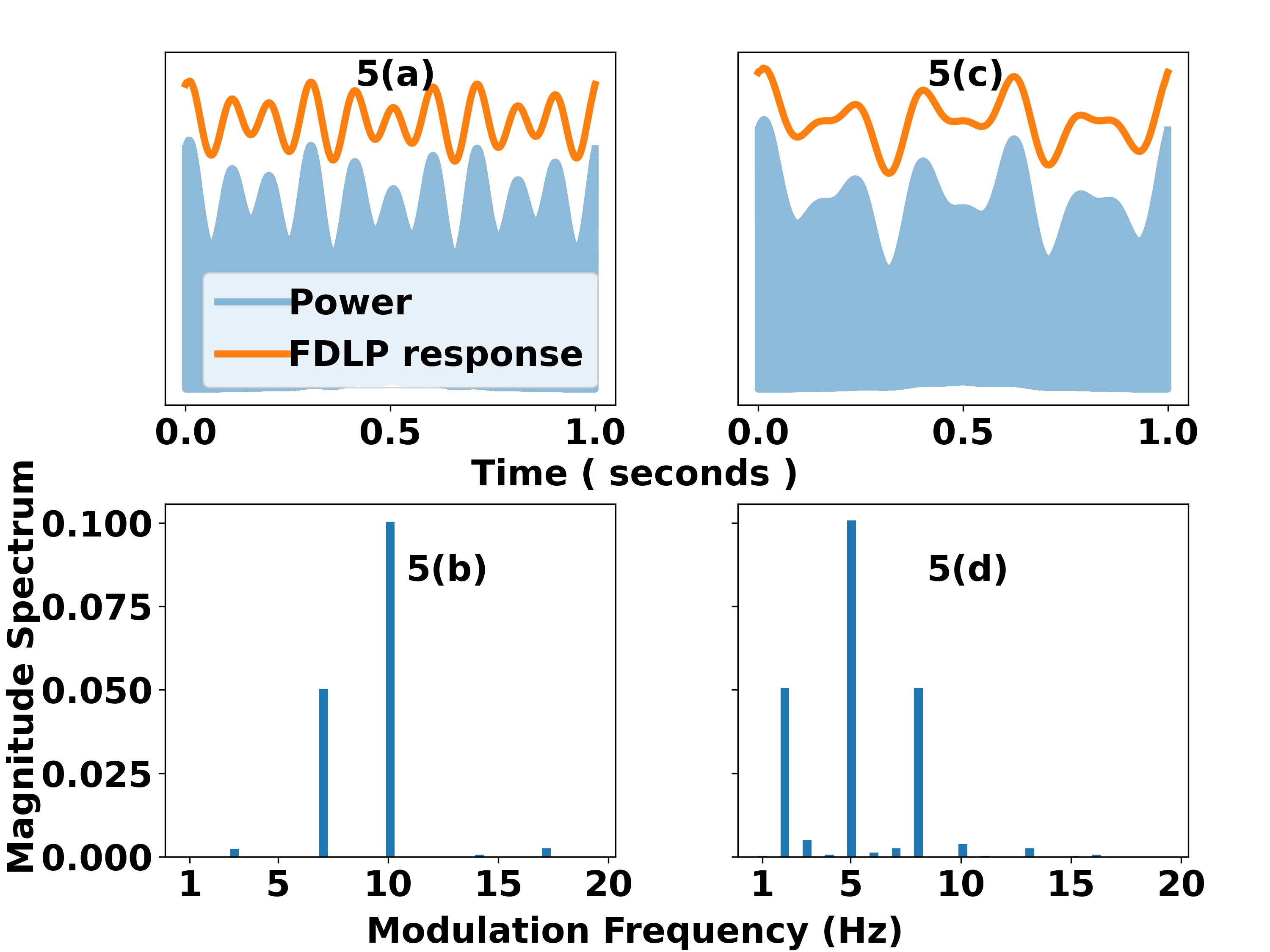}
    \caption{\\
    \textbf{Signal 1 (Two modulations)}: $N=2$, $f^{(m)}_1=7$ Hz, $f^{(m)}_2=10$ Hz,  modulation depths $\gamma_1=0.05, \gamma_2=0.1$, 5(a) shows the power of the signal and its FDLP response and 5(b) shows the FDLP derived modulation spectrum.  \\
    \textbf{Signal 2 (Three modulations)}: $N=3$, $f^{(m)}_1=2$ Hz, $f^{(m)}_2=5$ Hz,$f^{(m)}_3=8$ Hz, modulation depths $\gamma_1=0.05, \gamma_2=0.1, \gamma_1=0.05 $ and carrier frequency $f_c=1000$ Hz. 5(c) shows the power of the signal and its FDLP response and 5(d) shows the FDLP derived modulation spectrum. \\
    The modulation spectrum of both signals are independent of the phases of the modulations.}
    \label{fig:toy_example}
\end{figure}

With two simple examples, Figures \ref{fig:toy_example}(b) and  \ref{fig:toy_example}(d) show that for composite modulations of varying modulation depth, the complex FDLP processing is able to capture the true modulation spectrum. Minor harmonics are generated due to the non-linear log transformation of the approximated envelope.

\subsection{A Computational Benefit of Complex FDLP}
As is evident from Figure \ref{fig:compare_fits}, complex FDLP requires estimation of half the number of poles for the same approximation power - cutting the model order by half. On the contrary, computation of complex numbers is more resource intensive in comparison to real numbers - resulting in two conflicting influences on the total computational time. In an empirical study with 5000 examples of 1.5 second long speech signals using 300 order conventional FDLP and 150 order complex FDLP the observed average computation times are approximately $12.5$ and $10.5$ ms respectively on a 2 GHz Intel Core i5 processor - a $16\%$ reduction in computation time. 

FDLP based features, namely LP-TRAPS \cite{athineos2004lp}, modulation vectors \cite{sadhu2019m}, FDLP-spectrogram \cite{sadhu2021radically} require per frame cepstral computations in several frequency sub-bands over one utterance span and using complex FDLP yields significant computational speed-ups. In the following section we look at the ASR performance of FDLP-spectrogram derived with complex FDLP.

\subsection{FDLP-spectrograms \cite{sadhu2021radically}}
FDLP-spectrograms compute FDLP approximations to the power in different frequency sub-bands over 1.5 second Hanning windowed speech. Spectrograms of an arbitrary length speech signal is obtained by overlap-add of 1.5 second long spectrograms. Additionally these spectrograms are designed to preserve speech related modulations exploiting the FDLP machinery. Table \ref{tab:results} shows the results of End-to-end ASR models trained with \texttt{espnet2} recipes from ESPnet \cite{watanabe2018espnet} using complex FDLP based FDLP-spectrograms. 

\begin{table}[h]
\centering
\begin{tabular}{@{}lcc@{}}
\toprule
Data set & \begin{tabular}[c]{@{}c@{}}Mel-spectrogram \\ WER \%\end{tabular} & \begin{tabular}[c]{@{}c@{}}FDLP-spectrogram\\ WER \%\end{tabular} \\ \midrule
REVERB \cite{kinoshita2013reverb}  & 5.6                                                               & 4.8                                                               \\
CHiME 4 \cite{vincent2017analysis} &  12.6                                                                &     12.4                                                              \\ \bottomrule
\end{tabular}
\vspace{10pt}
\caption{ESPnet end-to-end ASR performances are shown for two different speech data sets. REVERB training data set consists of simulated reverberated speech and CHiME 4 consists of simulated noisy speech. Evaluation is done with real multi-channel reverberated and noisy speech data respectively. FDLP spectrograms are computed with 50 filters and FDLP model order 80 whereas Mel-spectrogram features are the default log-filterbank energy features computed by ESPnet. The ASR models are trained without speed perturbation and SpecAugment.}
\label{tab:results}
\end{table}
\section{Conclusions}
In this work, we propose a complex FDLP model that allows us to explicitly define and compute speech modulations from an all-pole approximation of the speech power trajectory. We show two-fold advantage of complex FDLP over conventional FDLP models, namely
\begin{itemize}
    \item An un-ambiguous interpretation of the magnitude of the complex cepstrum of the complex FDLP model as speech modulations - a feature that cannot be attributed to the conventional FDLP model.
    \item Improvement in computational time of all-pole model fits to the speech power envelope over conventional FDLP.
\end{itemize}
 In conclusion, we showed the end-to-end ASR performance of complex-FDLP based FDLP-spectrogram features on two speech data sets. 
\vspace{-10pt}
\section{Acknowledgements}
This work was supported  by the Center of Excellence in Human Language Technologies, The Johns Hopkins University and by a gift from Google Research.
\section{Appendix A}
\label{appendix:A}
Consider 
\begin{enumerate}
    \item A discrete time signal $x[n]$ with $N$ samples
    \item A Discrete Fourier Transform (DFT) pair \\
    $x[n] \overset{\mathcal{F}}{\longleftrightarrow} X[k]$
\end{enumerate} 
Then
\begin{eqnarray}
    &&|\mathcal{F}^{-1}( x[n] )|  \nonumber \\
    &=&\left|\frac{1}{\sqrt{N}}\sum_{n=0}^{N-1}x[n]e^{\frac{j2\pi kn}{N}}\right|  \nonumber \\
    &=& \frac{1}{\sqrt{N}} \left|\sum_{n=0}^{N-1}x[n]\cos{\frac{2\pi kn}{N}} + j\sum_{n=0}^{N-1}x[n]\sin{\frac{2\pi kn}{N}} \right|  \nonumber \\
    &=& \frac{1}{\sqrt{N}} \sqrt{ \left(\sum_{n=0}^{N-1}x[n]\cos{\frac{2\pi kn}{N}} \right)^2 + \left( \sum_{n=0}^{N-1}x[n]\sin{\frac{2\pi kn}{N}} \right)^2  }  \nonumber \\
    &=& \frac{1}{\sqrt{N}} \left|\sum_{n=0}^{N-1}x[n]\cos{\frac{2\pi kn}{N}} - j\sum_{n=0}^{N-1}x[n]\sin{\frac{2\pi kn}{N}} \right| \nonumber \\
    &=&\left|\frac{1}{\sqrt{N}}\sum_{n=0}^{N-1}x[n]e^{\frac{-j2\pi kn}{N}}\right| \nonumber \\
    &=&|\mathcal{F} (x[n])| \nonumber 
\end{eqnarray}

\bibliographystyle{IEEEtran}

\bibliography{mybib}


\end{document}